# Making statistical methods more useful: some suggestions from a case study

12 November 2012

*Michael Wood*
University of Portsmouth Business School
OPSM Department, Richmond Building
Portland Street, Portsmouth
PO1 3DE, UK
+44 (0) 23 92844168

michael.wood@port.ac.uk .

## Abstract

I present a critique of the methods used in a typical paper. This leads to three broad conclusions about the conventional use of statistical methods. First, results are often reported in an unnecessarily obscure manner. Second, the null hypothesis testing paradigm is deeply flawed: estimating the size of effects and citing confidence intervals or levels is usually better. Third, there are several issues, independent of the particular statistical concepts employed, which limit the value of *any* statistical approach – e.g. difficulties of generalizing to different contexts, and the weakness of some research in terms of the size of the effects found. The first two of these are easily remedied – I illustrate some of the possibilities by re-analyzing the data from the case study article – and the third means that in some contexts a statistical approach may not be worthwhile. My case study is a management paper, but similar problems arise in other social sciences.

# Making statistical methods more useful: some suggestions from a case study

## Introduction

Statistical methods are widely used in research. There is a vast amount of supporting theory, practical tips, examples of good practice, and so on, to support these methods. However, some fundamental aspects of the way statistical approaches are typically used sometimes seem problematic – even in studies published in respected journals whose reviewing process ensures that the obvious pitfalls are avoided. This paper considers three broad areas which are often problematic: the user-friendliness of the concepts used, the use of hypothesis testing, and issues about the usefulness of the general statistical approach which apply regardless of the particular methods employed. The paper proposes some possible ways of addressing some of these problematic aspects. It should be of interest to anyone concerned about the usefulness of statistical results – either as producers or consumers of statistical analysis.

My approach is to focus on a single, but typical, published research paper and to look at some of the problems with the analysis and presentation of the results, and at some alternative approaches. The case study paper is in a management research journal, but the issues raised are likely to be relevant to many other research projects in the social sciences. Obviously, no firmly generalizable conclusions are possible from a sample of one paper. However a case study approach like this, by analyzing an illustrative example in depth, can suggest *possibilities* which may – and very probably do – have a wider applicability. Ideally, I would analyze a representative sample of research studies, but the detail on which my argument depends makes this strategy impracticable.

I chose Glebbeek and Bax (2004) as my illustrative paper because it was published in a respected journal (the *Academy of Management Journal*), concerns a topic which can be appreciated without detailed knowledge of the literature, is clearly written, and the statistical approach used is fairly typical, involving regression and hypothesis testing. The aim is not to produce a critique of this paper, but to explore issues of wider concern for the use of statistics in research. I am very grateful to Dr. Arie Glebbeek for making the data available; this has enabled me to carry out some of the suggestions discussed below.

Glebbeek and Bax (2004) "tested the hypothesis that employee turnover and firm performance have an inverted U-shaped relationship: overly high or low turnover is harmful", with the optimum level of turnover lying somewhere in the middle. To do this, they analyzed data from "110 offices of a temporary employment agency" in the Netherlands. One of their analyses leads to Figure 1 below. The scattered points in Figure 1 each represent a single office, and the general pattern shows how performance ("net result per office" in Dutch guilders per full time employee per year in 1995 prices) varies with employee turnover. The solid line represents a best guess prediction for an office with a mean level of absenteeism (3.9%) and a mean age of staff (28.4 years) in one of the three regions of the study: the method used to make this prediction is discussed below. Glebbeek and Bax, 2004, mention but do not show this graph, although graphs of curvilinear models are shown in two subsequent articles on the same theme in the same journal – Shaw et al, 2005, and Siebert and Zubanov, 2009. They did several variations of this analysis – for example, they tried relating performance to current turnover and to turnover in the previous two years. However, for my purposes here I will focus on the data which is the basis of Figure 1.

**Figure 1: Results and curvilinear predictions for Region 1 and mean absenteeism and age**

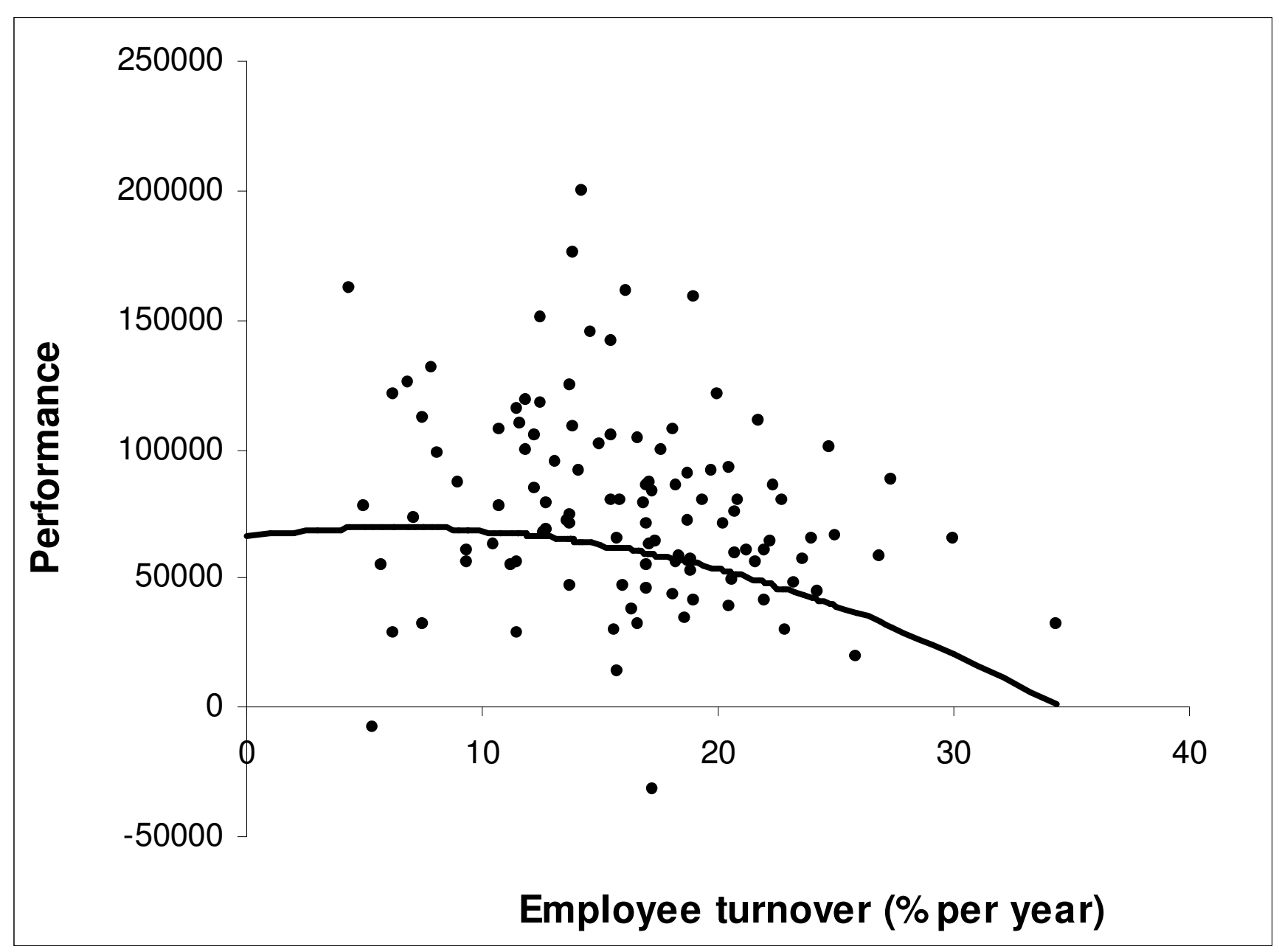

This graph, and the mathematical formulae on which it based, suggest that the optimal level of staff turnover is about 6%: for the best possible level of performance, 6% of staff would leave each year. Anything above or below 6% is likely to lead to poorer performance, and Figure 1 gives an indication of how much performance falls off – the prediction for performance is about 70,000 units if employee turnover is at the optimal level, but only about 3,000 if it is 34%. This information is of clear, practical interest to HR managers.

The analysis in Glebbeek and Bax (2004) used standard regression techniques which are summarized briefly in the next paragraph. The first issue I discuss below is that these methods are unnecessarily obscure – I would expect readers unacquainted with mathematical statistics to find some aspects of the next paragraph difficult. (Many papers in the management literature use far more complex statistical methods, so the task of rendering the analysis more transparent is more urgent, but also possibly more difficult. My aim here is simply to demonstrate the possibilities in a straightforward example.)

The regression models used “net result per office” (p. 281) as the dependent variable, staff turnover and the square of turnover as independent variables, and also three control variables. (Including a square term in the regression is a standard method of testing hypotheses about U-shaped relationships.) The results are presented, in the conventional way, by tables of standardized regression coefficients for the various models, supplemented by symbols to denote different ranges of *p* values (Tables 2 and 3 in Glebbeek and Bax, 2004). In all cases the coefficients were as predicted by the inverted U shape hypothesis: the regression coefficients for the (linear) turnover terms were positive, whereas for the squared turnover terms the coefficients were negative. However, none of the coefficients for the turnover terms were statistically significant, although three of the four coefficients for the squared terms were significant ($p$ < 5% in two cases and 10% in the third). The discussion in the article argues that this provides reasonable support for the inverted U-shape in the context of the employment agency in question, but it "was not observed with certainty" (277). The model in the first analysis table (A in Table 2, corresponding to Figure 1 above) gives the standardized regression coefficients for the turnover and turnover squared terms of the model as 0.17 and –0.45 respectively, but neither is significant ($p$ > 10%). There is nothing in the table of results to tell the reader about the 6% optimum level of turnover or how much difference departures from this figure make (although similar information is mentioned in the discussion).

My first task below is to explain how these results could be presented in a more user-friendly, but equally rigorous, form. Figure 1, which is *not* in Glebeek and Bax (2004), is a start, but it is possible to go further – as, for example, in Table 1 below.

My second aim is to review the hypothesis testing framework. Glebbeek and Bax's paper tests the hypothesis that the relationship is an inverted U-shape. There are several problems here. Most obviously, the graph in Figure 1 seems marginal as an inverted U-shape because the decline on the left hand side (low employee turnover) is very slight. It could just as plausibly be interpreted as a slightly curved declining relationship between the two variables. The hypothesis is a bit fuzzy which makes a clear test difficult.

As is usual in management research, Glebbeek and Bax test their hypothesis by means of null hypothesis tests and the resulting *p* values (both >10% for Figure 1). However, there are several very strong arguments – discussed below – against this approach. One alternative suggested here is to cite a confidence level for the hypothesis – this comes to only 65% (the source of this figure is explained below). This means that, on the basis of the data, we can be 65% confident that an inverted U-shape pattern would result if we analyzed *all* the data from similar situations. This seems far more useful than citing *p* values.

The data, and so the conclusions from any analysis, are based on one organization in one country in one era (the late 1990s): there is obviously no guarantee that a similar pattern would occur in other contexts. And, even given this, the scatter apparent in Figure 1 suggests that that staff turnover is just one of many factors affecting performance. These are among the more general issues that are relevant whatever statistical approach is taken to the analysis of the data: my third aim is to review these.

The medical research literature provides an instructive contrast to management. Ensuring that doctors without statistical training understand results accurately may be a matter of life and death, unlike the situation in management where most managers probably ignore most research. Null hypothesis testing is used much less in medicine and guidelines from journals (BMJ, 2011) and regulatory authorities (ICH, 1998) often insist on citing confidence intervals (to be discussed below). And the fact that the management environment is far less predictable than the human body studied by medical research also has implications for the way statistics are used.

The present paper analyses just one research study, and the detail of the analysis is clearly specific to this particular study. However, in the concluding section I derive some more general recommendations derived from this single example: these generalizations must, of course be tentative.

## Critiques of statistical methods as used in management research

Statistical analysis is of clear use for many tasks – e.g. modeling house prices, predicting which potential customers are most likely to buy something, and analyzing the results of experiments (Ayres, 2007). In examples like these the influence of noise variables may be substantial, but statistical methods enable us to peer through the fog and discern a tendency which is sufficiently reliable to be useful.

There is a large literature on the pros and cons of the different approaches to statistics (especially the Bayesian approach and how it compares with conventional alternatives), on the importance of particular methods and problems with their use (e.g. Becker, 2005; Cashen and Geiger, 2004; Vandenberg, 2002), on the importance and difficulties of educating users of statistics and readers of their conclusions, and, of course, on the derivation of new methods. However, there is surprisingly little by way of critique of statistical methods and their application in general terms.

One paper which does give such a critique of statistical modeling – in management science – is Mingers (2006). He claims that statistics, in practice, adopts "an impoverished and empiricist viewpoint", by which he means that it largely fails to go "beneath the surface to explain the mechanisms that give rise to empirically observable events". This is undoubtedly true in many contexts: Figure 1 indicates that Glebbeek and Bax's (2004) inverted U shape expresses a rather weak tendency which fails to incorporate the noise factors whose importance is clear from the scatter in Figure 1 (and the value of Adjusted $R^2$ which is 13%). Statistical analyses like this provide a partial, or probabilistic, explanation. If a satisfactory deterministic explanation is available, then a statistical model is not called for; in this sense, statistics is the method of last resort, but still a potentially useful approach when we do not fully understand what is happening.

One commonly reported problem with statistics is that many people – including some researchers and some of their readers – find the concepts and techniques difficult to understand. This is particularly true of null hypothesis testing – which is a convoluted concept involving trying to demonstrate "significance" by assuming the truth of a probably false null hypothesis. The obvious approach to dealing with problems of understanding is to call for more, and better, statistical education, and there is a very large literature, and several journals, on this topic.

An alternative approach to the education problem is to acknowledge that there are too many complicated techniques for researchers and readers to learn about (Simon, 1996, points out that people generally have about 10 years to become a specialist and this imposes a limit on the amount of expertise that can be mastered), so efforts should be made to present results that can be understood with the minimum level of technical understanding which is possible without sacrificing the rigor and usefulness of the analysis (Wood, 2002; Wood et al, 1998). This could be on the level of redefining output measures to make them more user-friendly, or using methods whose rationale is closer to common sense than conventional methods based on probability theory – this is one of the advantages of resampling methods such as bootstrapping (e.g. Diaconis and Efron, 1983; Simon, 1992; Wood et al, 1999; Wood, 2005). However, in practice, these opportunities are very rarely taken: one of my aims in this article to demonstrate some of the possibilities.

The user-friendliness issue is one, minor, strand in the debate about the pros and cons of different approaches to statistics. Another issue which deserves mention here is the debate about the role of null hypothesis significance testing (and *p* values). This is the standard method used in management, and most social sciences, of answering questions about how reliably we can generalize from a limited sample of data. There are however, very strong arguments, put forward in numerous books and articles over the years, against the use of these tests in many contexts (e.g. Cohen, 1994; Gardner and Altman, 1986; Kirk, 1996; Lindsay, 1995; Morrison and Henkel, 1970; Nickerson, 2000), and in favour of alternative approaches, such as the use of confidence intervals. According to Cohen (1994) "after four decades of severe criticism, the ritual of null hypothesis significance testing – mechanical dichotomous decision around a sacred .05 criterion – still persists." He goes on to refer to the "near universal misinterpretation" of *p* values. More recently, Coulson et al (2010) concluded on the basis of a survey of 330 authors of published papers that interpretation of *p* values was "generally poor" – and this is among authors, not readers. There is no space here for a general review of the arguments, but I will discuss the issues as they apply to Glebbeek and Bax (2004) in the next section.

Finally, it is important to note the obvious fact that there are alternatives to statistical methods. The simplest is to use case studies to illustrate and explore what is possible without any attempt to estimate population statistics (Yin, 2009, Wood and Christy, 1999). This is essentially the method I am adopting in this paper.

I turn now to a discussion of Glebbeek and Bax (2004). I will start with issues of user-friendliness, then progress to a discussion of the hypothesis testing approach which has been adopted, and finally consider the difficulties which would beset any statistical approach in this context.

## User-friendliness of the statistics in Glebbeek and Bax (2004)

This covers both the user-friendliness of the way the statistical concepts are described, and the user-friendliness of the concepts themselves. Readers may feel that readers of a technical research journal should be expected to understand technicalities without help. However, the extent of the expertise required means that it does seem reasonable to present results in as user-friendly manner as possible, provided this does not lead to the article being substantially longer, or to a sacrifice in the rigor and value of the analysis. As explained in the introduction, Glebbeek and Bax (2004) give the standardized regression coefficients for the turnover and turnover squared terms of the model as 0.17 and –0.45 respectively with $p > 10\%$ in both cases, and also standardized coefficients and *p* values for the three control variables. Figure 1 above is a step towards making this more user-friendly: Table 1 below gives a more detailed analysis.

**Table 1. User-friendly parameters for the model in Figure 1**

| | *Best estimate* |
|---|---|
| Location of optimum (annual employee turnover for best performance) | 6.3% |
| Inverted U-shape Curvature* | 86.7 |
| Predicted impact of 1% increase in absenteeism | -3,330 Guilders per FTE |
| Predicted impact of 1 year increase in average age | -831 Guilders per FTE |
| Predicted difference between neighboring regions (with Region 1 having the lowest performance) | 15,465 Guilders per FTE |
| Predicted maximum performance for Region 1, and mean absenteeism (3.8%) and mean age (28) | 69,575 Guilders per FTE |
| Estimated overall prediction accuracy (Adjusted $R^2$) | 13% |
| Confidence level for Inverted U-shape hypothesis | 65% |

*measures how curved the line is, with 0 representing a straight line, and larger numbers representing a more pronounced inverted U-shape curvature.

Only one of the figures in Glebbeek and Bax's (2004) Table 2 appears in Table 1: the value of adjusted $R^2$ (0.13) which I have rewritten as 13% to emphasize the fact that it can be regarded as a proportion. This statistic is an estimate of "the proportional reduction in error from the null model (with no explanatory variables) to current model" (King, 1986, p. 676), which I have summarized as "estimated overall prediction accuracy": 100% accuracy would imply that all predictions are completely accurate; 0% refers to a prediction making no use of the independent variables. Alternatively the phrase "proportion of variation explained" could be used but this seems less direct and the word "explained" may be misleading. The general point here is that it is a good idea to use labels which are as informative as possible; "adjusted $R^2$" is likely to convey nothing to the uninitiated.

The other statistics in Table 1 are all different from those given by Glebeek and Bax. Glebbeek and Bax give *standardized* regression coefficients which are awkward to interpret in a useful way (King, 1986); Table 1 gives the equivalent unstandardized coefficients for the control variables. For example, the standardized regression coefficient for the first control variable, absenteeism, is –0.19; in Table 1 the equivalent unstandardized coefficient ( –3,330) is given and its meaning described in terms of the scenario being modeled.

Instead of describing an inverted U-shape curve is in terms of standardized regression coefficients for the linear and squared terms (0.17 and –0.45), we could use unstandardized coefficients (1097 and –86.7) but these are still difficult to interpret. More usefully we could cite the coefficients in Table 1: the Location of the optimum (6.3%) and the Inverted U-shaped curvature (86.7). These are mathematically equivalent to the standard regression outputs presented by Glebbeek and Bax in the sense that the former can be calculated from the latter by simple formulae, and vice versa (Wood, 2012a). There is no loss of information, but it is in a format which makes it easier to relate to reality.

As an example, the performance of an office with an employee turnover rate 2% above the optimum (8.3%), in Region 1 with mean absenteeism and mean age would be predicted as:

$$69575 - 86.7(8.3\text{-}6.3)^2 = 69228$$

using the equations in Wood (2012a). In this equation the curvature obviously represents the extent to which performance falls as the employee turnover rate departs from its optimum value. The impact of the control variables can easily be added: if, say, absenteeism were 5% above the mean, then the predicted performance would be reduced by 5x3,330 to 52,578.

Finally there are no *p* values in Table 1. Instead there is a confidence level for the hypothesis. The derivation of this, and the reason for not giving *p* values are discussed in the next section.

## Problems with hypothesis testing, and suggested alternatives

Glebbeek and Bax "tested the hypothesis that employee turnover and firm performance have an inverted U-shaped relationship: overly high or low turnover is harmful". Formulating their research aim in terms of testing a hypothesis does, at first sight, give them a clear aim and a good headline to report in the literature. It is also conventional in research which seeks to be "scientific".

However, in this case, which is by no means unique, there are three obvious difficulties with the idea of testing this hypothesis:

1. The hypothesis is rather *fuzzy*. Figure 1 above is marginal as an inverted U-shape because the performance only just falls off as the turnover falls below the optimum (and the lack of data for low values of turnover means the evidence for this part of the line is weak). The scattered points in Figure 1 could plausibly be modeled by a straight line showing a declining trend: in practice these two possibilities merge into each other. The idea of testing a hypothesis probably derives its status from hypotheses like Einstein's $E=mc^2$; however, the inverted U-shape hypothesis here is far less impressive!

2. The hypothesis is rather *obvious*. If one imagines an organization where the employee turnover rate is over 100%, common sense suggests that performance is likely to be relatively poor. On the other hand, if the turnover were 0%, then this suggests that there is likely to be a lack of new ideas and energy, or that the organization is doing so badly that nobody can get another job. This means there must be an optimum level of turnover somewhere between the extremes, so the pattern *must* be an inverted U-shape.

3. Merely testing the hypothesis *ignores a lot of useful information.* Numerical information like the location of the optimum (6% for Figure 1), or how much difference departures from the optimum make, are irrelevant from the point of view of testing the hypothesis – which is a pity because these are likely to be what is most interesting in practice. It may be, for example, that in other sectors the optimum level of staff turnover is much higher – this is the sort of detail that is likely to be of interest to both theoreticians and practitioners.

These three points suggest that instead of testing a rather fuzzy and obvious hypothesis, a more useful aim for a research project like this is to *measure* things like the optimum level of employee turnover, and to *assess the shape of the relationship* between performance and employee turnover as illustrated by Figure 1.

## Problems with null hypothesis testing

The three arguments above concern the idea of hypothesis testing in general. As is conventional in management research, the particular approach used by Glebbeek and Bax to test the hypothesis is that of setting up a null hypothesis and then estimating the probability ($p$) that the data, or similarly extreme data (as measured by the test statistic), could have resulted from this null hypothesis. If this $p$ value is low we then conclude that the data is not consistent with the null hypothesis so this must be false, and an alternative hypothesis must be true.

Testing the inverted U-shape hypothesis like this is particularly problematic and will be discussed in the next section. Here we will consider the $p$ value for the predicted impact of turnover (i.e. the regression coefficient) in the linear (straight line) model for the data in Figure 1 which is less than 0.01; a more exact figure, using the Excel Regression Tool, is 0.007. This is worked out using the null hypothesis that employee turnover actually has no impact, positive or negative, on performance. The $p$ value indicates that chance fluctuations would lead to a value of -1,778 (the value actually observed) or less, or +1,778 or more, with a probability of 0.7%. This low probability means that the observed data is most unlikely to be a consequence of the null hypothesis, so we can assert that the evidence shows there is a real negative impact which would be likely to occur again if we took further samples.

This is a rather convoluted argument which people often fail to understand correctly. There are at least three problems from the user-friendliness perspective:

1. The focus for understanding *p* values is the null hypothesis, not the hypothesis of interest. Glebbeek and Bax do not even mention the null hypothesis, but this is the basis for the definition of *p* values.

2. The *stronger* the evidence for an impact of turnover on performance, the *lower* the *p* value is: as a measure of the strength of evidence the *p* value scale is an *inverse* one.

3. What users intuitively want is a measure of how probable a hypothesis is, and some indication of the nature and strength of the relationship between the two variables. Although *p* values do not answer either question, it is almost inevitable that some users will assume they do. This is not just a problem from the user-friendliness perspective, of course: a measure which fails to tell people what they want to know is not a good measure even if understood correctly.

Besides the user-friendliness issues there are a number of further problems with null hypothesis testing, one which are relevant here.

4. There may be problems choosing a sensible null hypothesis. To test their inverted U-shape hypothesis Glebbeek and Bax had *two* null hypotheses: that the population values of the linear and the square term regression coefficients were both zero. This means, in effect, that there is *no* consistent pattern, straight or curved, between the two variables. This is unsatisfactory because there is no obvious way of combining the two *p* values, and because this null hypothesis is far too strong if taken literally. Figure 1 shows a clear declining trend, so the null hypothesis is fairly obviously false, but this does not mean the curvilinear hypothesis is true. Null hypothesis tests can effectively rule out the null hypothesis, but this is not helpful to provide evidence for an alternative hypothesis if there is more than one such hypothesis. As we saw above, the *p* value for the linear (straight line) model was 0.7%. This is based on the null hypothesis that increasing staff turnover has a *no* impact on performance. However, again, this is so unlikely that getting evidence that it is false is not really of interest. In both cases, the obvious null hypotheses used by Glebbeek and Bax do not deliver much interesting information.

## Measuring the size of the impact and using confidence intervals

One recommendation with the potential to deal with *all* these problems is to estimate the size and nature of the impact of employee turnover on performance, and then to express the uncertainty in this estimate by means of confidence intervals. We will start by discussing a linear model (Model 3 in Glebbeek and Bax, 2004) because this is more straightforward. According to the linear model the best estimate for the impact of an extra 1% staff turnover on performance is –1778 (the unstandardized regression coefficient). This is an estimate of a numerical quantity, does not involve any hypothesis, and avoids the problems of focusing on a fuzzy, obvious and distracting null hypothesis.

However this does not deal with the problem of sampling error: different samples are likely to produce different results, and it is unlikely that the sample result is exactly correct for the whole population. Null hypothesis tests provide one, unsatisfactory, way of approaching this problem; confidence intervals are often recommended as an alternative (e.g. Gardner and Altman, 1986, writing in the *British Medical Journal,* Cashen and Geiger (2004), and Cortina and Folger, 1998).

In Table 2 below, the best estimate for the impact of staff turnover is that each extra 1% will reduce performance by 1,778. However, the exact amount is uncertain: the confidence interval suggests that the true impact is somewhere between a reduction of 495 units and 3,060 units with 95% confidence. This interval *excludes* zero, which means that the significance level must be less than 5% (100% - 95%); in fact *p* is less than 1%, meaning that the 99% confidence interval would also include only negative values. On the other hand, the 95% confidence interval for the impact of age includes both positive and negative values, which means that it is not possible to reject the null hypothesis that age has no impact at the 5% significance level.

**Table 2. Confidence intervals for linear Model (3 in Panel A of Table 2 in Glebbeek and Bax, 2004)**

| | *Best estimate* | *Lower limit of 95% CI* | *Upper limit of 95% CI* |
|---|---|---|---|
| Predicted impact of 1% increase in staff turnover | -1,778 | -3,060 | -495 |
| Predicted impact of 1% increase in absenteeism | -3,389 | -6,767 | -10 |
| Predicted impact of 1 year increase in average age | -731 | -3,716 | 2,254 |
| Predicted difference between neighbouring regions (with Region 1 having the lowest performance) | 15,066 | 5,607 | 24,525 |
| Estimated overall prediction accuracy (Adjusted $R^2$) | 12% | | |

Presenting the predicted impacts and confidence intervals as in Table 2 avoids any mention of null hypotheses with their associated problems. It focuses on the relationship of interest rather than a hypothetical and almost certainly false hypothesis, and the confidence interval expresses the uncertainty in a far more transparent way than the *p* value. As we have seen, all the information provided by the *p* values can be derived from the confidence intervals, but the confidence intervals also give a lot of extra information.

Despite their advantages, confidence intervals are very rarely cited in management research. The situation is very different in medicine: confidence intervals are widely reported and recommended by journals (e.g. BMJ, 2011) and regulatory authorities (e.g. ICH, 1998).

## Confidence levels for hypotheses

Unfortunately, this approach is not so easy for assessing the confidence in the conclusion that the curve is an inverted U-shape because this is measured by two parameters – location and curvature in Table1. (The location is relevant to the existence of an inverted U-shape because if the optimum occurs for a negative value of turnover, then there will not be an inverted-U in the positive, meaningful part of the graph.) We could produce confidence intervals for the curvature and the location of the optimum turnover in Table 1, but the fact that there are two quantities here would make these unwieldy and difficult to interpret. So I next consider how we might apply the idea of confidence to a hypothesis.

Bootstrapping provides an easy approach to this problem. The idea of bootstrapping is to use the sample of data to generate “resamples” which mimic other samples from the same source. A group of such resamples can then be used to see how variable different samples are likely to be and so how confident we can be about the hypothesis. In the present case Figure 2 shows the prediction curves generated from four such resamples, and also the prediction from the original data (as in Figure 1).

**Figure 2: Predictions from data (bold) and 4 resamples for the model in Figure 1**

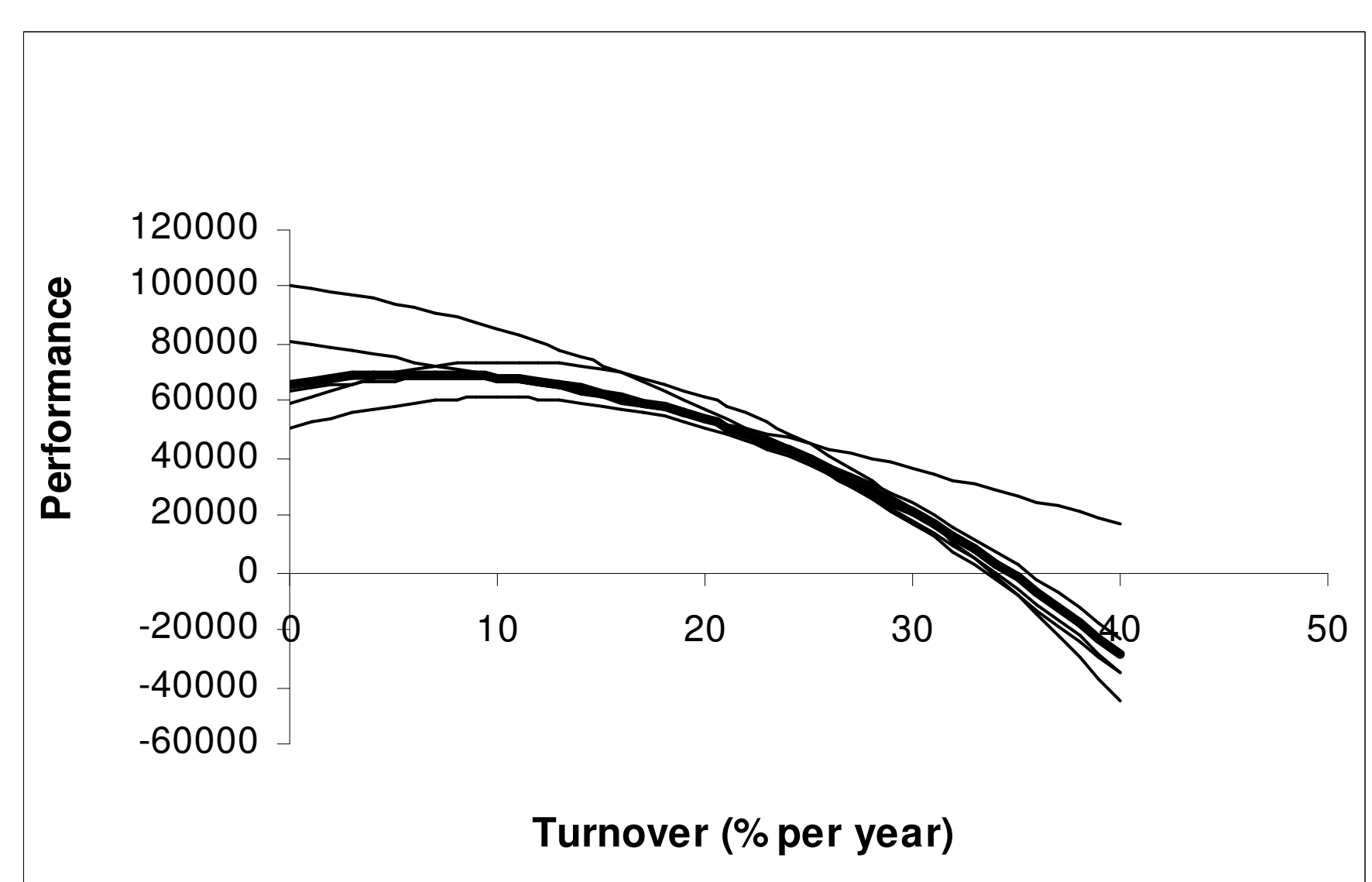

Figure 2 gives a clear demonstration of the fact that Figure 1 may be misleading, simply because two of the five lines are not inverted U shapes. With 10,000 resamples, 65% produced an inverted U shape (with a negative curvature and positive value for the location of the optimum). *This suggests a confidence level for the inverted U shape hypothesis of 65%.*

There is a slightly more detailed explanation of the bootstrapping procedure in the Appendix. There is also an extensive literature on bootstrapping: there are simple explanations in Diaconis and Efron (1983), Simon (1992) and Wood (2005), and a more detailed explanation of the procedure used here in Wood (2012a).

However, unlike the confidence interval in Table 2, simply stating a confidence level of 65% for the inverted U-shape hypothesis gives little indication of how steep the curve is or what the optimum employee turnover is. The hypothesis does not distinguish between marginal curves and strongly curved ones.

We can deal with this problem to some extent by assessing a confidence level for a stronger hypothesis. For example, we might insist that for a reasonable inverted U-shape the graph needs to drop at least 10,000 units on the left hand side – the confidence level in this case comes to 40%. However, the cut-off chosen is arbitrary because hypotheses like this are inevitably fuzzy.

We should also note that, strictly, a confidence level for an interval or a hypothesis is *not* the same as a probability of the truth of the hypothesis or of the true parameter value being in an interval (Nickerson, 2000: 278-280). Like null hypothesis tests, confidence intervals are based on probabilities of sample data *given* the truth about a parameter. To reverse these probabilities and find the probability of a hypothesis *given* sample data, we need to use Bayes' theorem and take account of prior probabilities. However, for many parameters, including the slope of a regression line and the difference of two means, the Bayesian equivalent of a confidence interval, the credible interval, is identical to the conventional confidence interval (Bolstad, 2004: 214-5, 247; Bayarri and Berger, 2004) provided we use "flat" priors (i.e. we assume a uniform prior probability distribution) for the Bayesian analysis. *This means that it is often reasonable to interpret confidence intervals and levels in terms of probabilities*: the only loss (from the Bayesian perspective) is that any prior information is not incorporated.

## General issues about the usefulness of the statistical approach

Let's suppose now that the recommendations above have been taken on board: results are presented in as user-friendly a manner as possible; confidence intervals are used whenever possible, and when not possible confidence levels instead of *p* values are used to quantify uncertainty. The statistical methods have been tweaked in accordance with the discussion above, but there are still important questions about the usefulness of the statistical methods in general terms – these are relevant regardless of the particular methods and concepts used.

Conclusions from statistical research are not deterministic, but are qualified by probabilities, averages or equivalent concepts. The relationship between employee turnover and performance is likely to be too complex for a complete, deterministic explanation of all the variables and their exact effects: the statistical approach is therefore worth considering in the absence of anything better. The following issues are relevant to questions about the value of a statistical approach.

## The strength or weakness of statistical results

Figure 1 shows a fairly weak U-shape in the sense that the fall-off on the left hand side is slight. The estimated accuracy of the prediction (adjusted $R^2$) is only 13%, as is obvious in rough terms from the scatter in Figure 1. And, the level of confidence that this shape is a feature of the underlying populations and would be repeated in further samples is fairly low – 65%. The results are weak on all three of these dimensions – *the strength of the effect* (how definite the inverted U-shape is)*, the consistency of the effect* (it is obvious that there are other factors besides employee turnover which are responsible for good or bad performance), and *the confidence level for the hypothesis*. Furthermore, as discussed in Glebbeek and Bax (2004), the prevailing wisdom is that performance has a tendency to fall as employee turnover rises, although common sense, for the reasons given above, suggests that an inverted U-shape of some kind is almost inevitable. For all these reasons Figure 1, and the data and analysis behind it, seems to add little value to what is already known.

## The nature and generality of the target context

If null hypothesis significance tests, or confidence intervals, are to be used correctly to analyze Glebbeek and Bax's data, we must assume that the sample used is a random sample from a specified target population. In practice, the sample used by Glebbeek and Bax was a *convenience sample* – the data concerned *all* branches of the employment agency in question which was chosen simply because it was available. At first sight there is no target population beyond the sample, Figure 1 is exact in relation to this sample, and there is no uncertainty due to sampling error. So what sense can we make of the 65% confidence level or the *p* values?

If we took another, similar organization with the same forces at work, or the same organization at a different time, we would be most unlikely to get Figure 1 exactly. A multiplicity of noise factors would mean that the next sample would be different – perhaps similar to one of the four resamples in Figure 2 above. We need to know how variable samples are likely to be due to these random factors, so that we can assess confidence levels for conclusions.

The standard terminology of populations is a little awkward here, so I will use the phrase *target context* to refer to the context the research is targeting, from which the sample can reasonably be considered a random sample, and to which the results can reasonably be generalized. In the absence of a formal sampling process this notion is inevitably rather hazy. (The target population would be a hypothetical population of offices "similar" to those in the sample, but this seems distinctly difficult to visualize.)

The nature of this target context deserves careful consideration. Glebbeek and Bax's results are based on data from a single organization in a single country (the Netherlands) in one time period (1995-1998), so perhaps the target context should be similar organizations in the same country at a similar time in history? Obviously, a different context may lead to a different pattern of relationship between staff turnover and performance, so their conclusions are qualified by words such as "can". The inverted U shape hypothesis is in no sense demonstrated in general terms, as Glebbeek and Bax acknowledge, but they have shown that it is a possibility because it applies to this one target context.

The scope of the target context is a key issue in this, and most other empirical management research. The difficulty with making the target context too broad is that it becomes difficult to obtain reasonable samples, and the specific contextual factors are likely to add to the noise factors making it difficult to get clear results. On the other side, if the context is too narrow this may lead many to conclude that the research is of little relevance.

The notion of a target context becomes more subtle when we consider the time dimension, or when we extend the idea to include what is possible. Most management research, and Glebbeek and Bax's is no exception, has the ultimate purpose of improving some aspect of management in the future. The aim of empirical research is to try and ascertain what works and what does not work. Let's imagine, for the sake of argument, that we had a similar data from a representative sample of a broader target context: all large organizations in Europe over the last ten years. This would certainly be useful, but the context might change over the next few years so we would have to be cautious in generalizing to the future. The difficulty with almost any target context for statistical management research is that it depends crucially on contextual factors which may change in the future. Although it is perhaps a worthy aim to extend our theories to incorporate these contextual factors, this may make the resulting theories messy and unmanageable. Perhaps we should try to focus on the core, immutable, truths instead? The difficulty, of course, is that there may be no such immutable truth other than the fact that it varies from situation to situation – in which cases statistical analysis would be only of limited interest.

A comparison with medical research is instructive. The target context here might be people, perhaps of a particular age or gender. With a management study the target context would typically be organizations or people in a particular business context. The problem with the business context, but not the medical one, is that it is an artificial context which may differ radically between different places or different time periods, making extrapolation from one context to another very difficult. Can conclusions about how staff turnover affects one employment agency's performance in an economic boom in the Netherlands be assumed to apply to universities England in the next century, or to social networking websites in California? Almost certainly not. On the other hand, in medicine, research with a restricted local sample may be of wider value simply because people are less variable from a medical point of view than business environments are from a management point of view.

### The necessity to use easily measurable variables

Statistical research has to focus on easily measurable variables. Otherwise it is not practical to obtain useful sample sizes. In the present case, employee turnover, performance, and the control variables absenteeism level, age and region are all easily measurable and available. Obviously, there are likely to be softer variables, which could not be so easily defined or collected, which may have an important bearing on performance.

### Alternatives to the statistical approach

Finally, we must remember other approaches, either as alternatives or as additions to, the statistical approach. Most obviously, case studies and "qualitative" research, which "can provide thick, detailed descriptions in real-life contexts" (Gephart, 2004: 455), might illuminate *how* high turnover influences performance, or give information about particularly interesting scenarios – perhaps even black swans (Taleb, 2008) – which would be not come to light via predictions like Figure 1 about what happens on average. This is not an argument against using a statistical analysis, but it may be an argument for supplementing it with a more detailed qualitative study of a smaller sample. This mixed methods principle seems to be widely accepted in theory, although not always in practice.

## Conclusions and recommendations

I have looked at three sets of problems concerning the statistical analysis in my case study paper. The first concerns user-friendliness: this can be improved by using more appropriate names for concepts (e.g. estimated overall prediction accuracy instead of adjusted $R^2$), and by changing the parameters cited themselves (e.g. the location of the optimum and the degree of inverted curvature instead of the regression coefficients for the linear and square terms). Some more possibilities are illustrated by Tables 1 and 2. Authors generally (but not always) include this sort of information in their discussion, but my suggestion is that the information given in tables of results should be in a more user-friendly form than is conventional – see Tables 1 and 2 above. There is no loss of information or rigor in doing this: it is not a matter of "dumbing down" but rather enhancing the accessibility of the research, and increasing the chance that results will be interpreted correctly by as many readers as possible. I have used my case study paper for illustration: some of the suggestions would carry over directly to other research, but my main aim is to establish the principle.

The second issue concerns hypothesis testing. Statistical research papers do *not* need lists of hypotheses – whose truth or falsity is often entirely obvious – to test. A more sensible aim is to assess the relationship between variables – as numerical statistics and/or in the form of graphs. Conventional quantitative research based on hypothesis tests is often strangely non-quantitative because readers are told very little about the *size* of impacts, differences or relationships. And instead of using the convoluted, uninformative and widely misinterpreted, *p* values, uncertainties due to sampling error can often be expressed as confidence intervals. In the example we looked at above, this resolved all the identified difficulties with null hypothesis tests.

Despite this, sometimes there may be reasons for testing a hypothesis. The truth or otherwise of the inverted U-shape hypothesis cannot easily be summed up by means of a single number: there is no obvious measure of “inverted U-shapeness”, so there is little choice but to formulate the research aims in terms of testing a hypothesis. My suggestion is that instead of trying to use *p* values to assess the strength of the evidence for this hypothesis, instead we use a *confidence level*. For Figure 1 this comes to 65%. We could also make the hypothesis a bit stronger as explained above – the confidence level for the stronger hypothesis is 40%. These confidence levels are far more straightforward and user-friendly than the conventional *p* values.

Third, we need to consider the value of statistical methods in general – issues to consider are the “strength” of the statistical results, the nature of the target context to which results can be generalized, and the extent to which the necessity to use easily measured variables distorts the research. The advantage of statistical methods is that they enable us to peer through the fog of noise variables to see patterns such as the curve in Figure 1. But Figure 1 illustrates the fuzziness of many statistical hypotheses: it is marginal as an inverted U-shape. And the fact that the data is a limited convenience sample means that it is difficult to generalize the conclusions to other organizations, times and places.

These conclusions and suggestions are based on a single research study. Obviously no firm conclusions can be drawn about how typical some of the problems described are, and the detailed suggestions made just apply to Glebbeek and Bax (2004). The bootstrap method for assessing the confidence level in the hypothesis is useful in this case, but for other hypotheses concerning, for example, the equality or difference of two means, simpler methods of assessing confidence levels may be feasible (Wood, 2012b). In the example above I have assessed the accuracy of the model as a percentage accuracy (adjusted $R^2$), but in other studies where there is an interest in making predictions it may be more sensible to give an estimate of the average error in a prediction from the model (e.g. the standard error). The statistics which are useful depend on the context and purpose of the research.

However, there seems to me little doubt that many of the issues highlighted above are widespread, so similar comments and suggestions are likely to be applicable to many other papers. (For example, ten of the eleven research articles in the September 2010 issue of the British Journal of Management presented *p* values, and eight were also organized round formal hypotheses.) Little effort is generally made to present results in an easily understood format.

Results are often cited as confirmation or otherwise of hypotheses, which are often fuzzy or obvious, with little or no discussion of how big the impacts or effects are. And statistical results are often fairly weak (although this may be disguised by the use of *p* values and large samples), and may be based on samples which make it difficult to extrapolate results credibly to the different environments which are likely in the future. The underlying idea of using statistical approaches in some contexts may be worth challenging.

## Acknowledgements

I am grateful to Dr Arie Glebbeek for making his data available to me, and to two anonymous referees for their valuable suggestions.

## Appendix: the bootstrap method for deriving confidence intervals and levels

There are 110 records in the data on staff turnover. The bootstrap method uses resampling with replacement: this means that we choose one of these records at random, then replace it so that we are starting again with the original sample, and then choose another at random, and so on until we have a "resample" of 110. All records in the resample come from the sample, but some records in the sample may appear several times in the resample, and others not all. We then follow this procedure repeatedly to generate multiple resamples.

Now, imagine that the population from which the real sample is drawn follows the same pattern as the sample. This means that 0.91% (=1/110) of the population will be like the first record in the sample, 0.91% like the second, and so on. This, in turn, means that to take a random sample from this population we want to choose a record like the first member of the sample with a probability of 0.91%, and similarly for the second, third and so on. But this is exactly what resampling with replacement achieves, so these resamples can be regarded as random samples from this imaginary population. This is not the real population, but it seems reasonable to assume it is similar, so that the variation between the resamples gives a good idea of sampling error when sampling a real population. In practice, experience shows that bootstrapping generally gives very similar results to conventional methods where these are possible. More details, and a link to the software used to derive the results above, are given in Wood (2012).